\documentclass[runningheads]{llncs}
\usepackage{graphicx}
\graphicspath{ {images/} }

\begin{document}
\title{Building an Ethereum-Based Decentralized Vehicle Rental System}

\author{Néstor García-Moreno\inst{1} \and
Pino Caballero-Gil\inst{1}\orcidID{0000-0002-0859-5876} \and Cándido Caballero-Gil\inst{1}\orcidID{0000-0002-6910-6538} \and
Jezabel Molina-Gil\inst{1}}
\authorrunning{N. García-Moreno et al.}

\institute{Department of Computer Engineering and Systems \and University of La Laguna. \and Tenerife. Spain. \and
\email{\{ngarciam, pcaballe, ccabgil, jmmolina\}@ull.edu.es}\\
\url{https://cryptull.webs.ull.es/}}
\maketitle              
\begin{abstract}

Blockchain technology, beyond cryptocurrencies, is called to be the new information exchange ecosystem  due  to its unique properties, such as immutability and transparency.
The main objective of this work is to introduce the design
of a decentralized rental system, which leverages 
smart contracts and the Ethereum public blockchain.
The work started from an exhaustive investigation on the Ethereum platform,  emphasizing the aspect of cryptography and all the technology behind this platform.
In order to test the proposed scheme in a realistic use, the implementation of a  web application for the rental of vehicles has been carried out.
The application covers the entire vehicle rental process offered in traditional web applications, adding more autonomy and ease of use to users. Following Ethereum application development guidelines, all business logic is located in the smart contracts implemented in the Ethereum network, where these contracts control the entire vehicle rental system of customers. While this is a work in progress, the results obtained in the first proof of concept have been very promising.

\end{abstract}

\section{Introduction}

\subsection{Background}

Until  recently, all electronic transactions between two parties have required centralized platforms, such as banks or credit card entities, in order to be able to mediate the interests of the transmitter and acquirer, and enable valid secure payments.
These platforms store the description of the items purchased and their price, and customers must interact with those platforms to purchase any item.
A feature of these platforms is that they all require a Trusted Third Party (TTP) to operate, resulting in many disadvantages for consumers. For example, consumers should usually register on each platform separately, share their private data with the owners of the platform, pay transaction fees, and depend on the security of the TTP.

A solution to overcome all these problems is given with the concept of Smart Contract (SC), which was first introduced by \cite{Szabo} as a digital protocol that facilitates an agreement process between different parties without intermediaries, enforcing certain predefined rules that incorporate contractual clauses directly in hardware and software.

The most important novelty that smart contracts include is the fact that each contract is executed on the nodes of a network and can be developed by anyone because a SC is a program that seals an agreement between two or more entities without the need for a TTP. In particular, every SC consists of a series of input variables, some output variables and a contract condition so that it is executed when the condition is met and the output variables are delivered to the entity indicated in the contract condition.

Before the appearance of the blockchain technology, there was no platform that could make SC a reality. Bitcoin  is an example of a specific SC, and Ethereum is one of the  platforms that allow to create SC in  general.
Ethereum is a set of  network, platform and protocol that shares many of basic concepts with Bitcoin, such as transactions, consensus and Proof of Work (PoW) algorithm.  Ethereum started off on the basis of PoW protocol to mine using  the computational brute force of the node. Now it is under the process of being moved to Proof-of-Stake (PoS) as the new basis of the distributed consensus algorithm to mine ethers by requesting tests of possession of such coins. With the PoS mechanism, the probability of finding a block of transactions, and receiving the corresponding prize, is directly proportional to the amount of coins that have been accumulated.

These characteristics of Ethereum allow the development of Decentralized Applications (DApps) without a server, thanks to smart contracts.
The use of smart contracts provides a layer of security and transparency to classic centralized applications. In the proposal here described, sensitive information is stored in the blockchain, preventing it from being modified or manipulated.

On the one hand, hash functions are used in this type of schema because they provide a random value of fixed length from an arbitrary input, and this process is not computationally reversible, so given a hash value it is practically impossible to obtain the original data, provided that the chosen hash function is robust. Hash functions are used to verify data integrity, in order to check whether they have been modified or not, since the corresponding associated hash value changes if the input changes. Therefore, they are used in blockchain because it is a way to verify whether the information stored in each block has changed or not. Bitcoin uses the SHA256 hash function, which always returns a string of 64 hexadecimal characters, that is, 256 bits of information.  
On the other hand, public key cryptography is used in the transactions of decentralized networks.  In this type of cryptography, each user has both a public key accessible to anyone, and a private key used to sign transactions, and that  must be kept secret. In asymmetric cryptography the private key is used to decrypt and the public key to encrypt messages.

The structure of this paper is as follows. In Section 2, some works related to the proposal are mentioned. In Section 3, the decentralized applications ecosystem is described. In Section 4, the proposed user application is detailed, including  technology,  contracts and  implementation. Section 5 includes a brief security analysis. Finally, the paper is closed with the conclusions and future works in Section 6.

\section{Related Works}

In this section several  research publications related to different DApps for vehicle rental are mentioned. The methodology for conducting the review is based on a synthesis of several comprehensive literature reviews on sources like Google Scholar and electronic databases. On the one hand, search queries in Google Scholar were  created using the following keywords: "blockchain", "decentralized applications"  and "vehicle rental".  On the other hand, electronic databases were used to find out features of Peer-to-Peer car rental companies.

Although the blockchain technology is a relatively new concept in the field of Information Technology, some reviews on several related concepts have already been published.
The first book on the market that teaches Ethereum and Solidity was \cite{Dannen}.
The  publication \cite{Vujicic} gives a brief introduction to blockchain technology, bitcoin, and Ethereum.
The work \cite{Lesavre} categorizes blockchain-based Identity Management Systems into a taxonomy based on differences in blockchain architectures, governance models, and other features. 

With regard to  specific blockchain-based applications for objectives similar to  this work, several authors have described some proposals. 
A DApp  for the sharing of everyday objects based on a SC on the
Ethereum blockchain is demonstrated in the paper \cite{Bogner}.
The work \cite{Huh} proposes using blockchain to build an Internet of Things system.
The paper \cite{Niya} introduces the design and implementation of an Android-based Peer-to-Peer purchase and rental application, which leverages Smart Contracts and Ethereum public blockchain. 
The work \cite{Ren} presents a car rental alliance solution based on internet of vehicles technology and blockchain technology.

An investigation of existing DApps reveals that only a few exploit SC to develop applications for the purpose of  flexible, valid and secure transaction executions in the case of rental use.
HireGo \cite{HireGo} is a decentralized platform for sharing cars that has been operating since 2019. Its schedule highlights the launch of its own token to use its service and the implementation of the Internet of Things and smart cars to automate operations with its token. It is the first DApp in the Ethereum Test Network (Test Net) to share vehicles in the United Kingdom.
Darenta \cite{Darenta} is a Peer-to-Peer car rental market that connects people who need to rent a private car with vehicle owners.
Helbiz \cite{Helbiz} is a decentralized mobile application to rent bicycles and electric scooters.

The conclusion of this literature review is that   practical applications of blockchain using Ethereum to develop DApps is a field that is beginning to be explored and on which there is still much to study and improve. 

In particular, the area of application of e-commerce for purchase and rental is one that has  the greatest potential, although before several problems, such as efficiency and privacy protection, have to be solved. 
Some common problems in online car  renting marketplaces such as Avis, Enterprise or Hertz are the need for an online platform operator to act as TTP, the lack of privacy of users when using those platforms, and the need for individual repetitive registration on each platform. In this paper we propose that these problems can be solved using a Ethereum DApp for rental cars that replaces the intermediary.
That is just the main goal of this work.

\section{DApp Ecosystem}

The first logical step to build  a DApp is to set up an environment that allows the development of smart contracts. This section details some of the technologies and software used to develop in Ethereum the contracts proposed in this work.

In a typical application development environment, one has to write code and then compile and execute the code on their own computer. However, the development of DApps is different because it brings decentralization to code execution, allowing any machine on the network to execute your code for you. If you want to develop a DApp without paying costs for the implementation and execution of functions within the contracts you are testing, you have to solve the challenge presented at this point.

The best solution for this problem is to use a blockchain simulator, which is a lightweight program that contains implementations of the Ethereum blockchain to be ran locally with minor modifications, such as control over mining speeds. As such, it is possible to mine blocks instantly and run decentralized applications very quickly.

Below is a review of several of the most popular technologies that allow the realization of this development cycle using a blockchain simulator:
\begin{itemize}
\item \textbf{Truffle}: \cite{Truffle} Framework that encompasses the entire Ethereum ecosystem, including decentralized application development, application testing and deployment on the Ethereum blockchain. It allows the development, compilation, test and deployment of smart contracts in the blockchain. It also allows the maintenance of private networks or public test networks (Rinkeby, Kovan, Ropsten). In addition, this framework contains an integrated command line interface for application development and configurable scripts to automatically launch contracts to the blockchain. 
\item \textbf{Web3}: \cite{Web3} Collection of libraries which allow  interaction with a local or remote Ethereum node using a HTTP or IPC connection. 
\item \textbf{Solidity}: \cite{Solidity} It is an object-oriented, high-level language for implementing smart contracts. Solidity was influenced by C++, Python and JavaScript and  is statically typed, supports inheritance, libraries and complex user-defined types among other features. 
\item \textbf{Ganache/TestRPC}: \cite{Ganache} Environment for the deployment of private or local blockchains for testing and preproduction. It is the old TestRPC, now in disuse. It comes integrated with Truffle. This environment allows  to inspect the entire record of the transactions of the blocks and the whole chain of  private blockchain. 
\item \textbf{Embark}: \cite{Embark}
Framework to develop and implement decentralized applications without a server. Embark is currently integrated with  Ethereum Virtual Machine (EVM),  InterPlanetary File System (IPFS) and decentralized communication platforms. 
\item \textbf{Serpent}:  \cite{Serpent} Smart contract-oriented programming language inspired by Python. It is currently in disuse.
\item \textbf{Metamask}: \cite{Metamask} Browser extension that acts as a bridge between the blockchain and the browser. It allows  to visit and interact with DApps without running a full Ethereum node. 

\end{itemize}

\section{User Application}

AutoRent is the solution presented in this paper.
It is a decentralized web platform for car rental using smart contracts in the Ethereum blockchain. It has been developed using the JavaScript Truffle framework, which is one of the most popular in the development of DApps. This framework includes the entire development cycle of an application: preproduction, production, and deployment.

For the development of smart contracts, Solidity, a high-level language oriented to contracts, has been used. Its syntax is similar to that of JavaScript and is specifically focused on the EVM.

\subsection{Smart Contracts}

Truffle includes a compiler for smart contracts developed in Solidity, with the tools necessary for deployment on the Ethereum network (migrations).
It has a Command Line Interface (CLI) to facilitate this task. This interface allows  to compile, deploy (migrate), tests, etc.

\subsubsection{Driving License Contract} 

One of the main problems of developing a decentralized vehicle rental application is to verify the authenticity of the driver's license.
The driving license in Spain coincides alphanumerically with the National Identity Document (DNI), but for the driving license there is no official platform of the Spanish Government to confirm its authenticity, so the lessee could provide a false permit or not have it when renting a car. 

The ideal way to remedy this security breach would be for the Government to have an official platform (which may be a blockchain) where to consult whether a person has a driving license. 

As an alternative, a private blockchain has been created in this work, in which  the platform administrator in charge of the car rental company can introduce the driving  licenses that have been previously validated, giving the corresponding clients  access to rent  vehicles.
This contract is simply composed of a method to introduce previously validated driving licenses, and another method that lists the driving licenses that have been introduced.

\subsubsection{RentACar Contract} 
In order to lease a vehicle belonging to a rental  fleet, a smart contract is necessary to verify two different aspects: on the one hand,  that the desired vehicle  exists in the private blockchain of the company, and on the other hand, that it is not already rented. If everything is correct, this contract  establishes a mapping between vehicle and customer  data, declaring it as a lessee.
Each day that passes, the contract automatically removes the daily price of the rented vehicle from the customer's deposit. If the deposit is insufficient or non-existent,  an extra expense   will be charged. At any time, the client can add funds  to the contract in order to extend the rent or to avoid surcharges. In order to return the car, a user must not have pending charges. Otherwise, the transaction will be rejected and he/she will be prompted to add the funds he/she  owes.

Once the vehicle is returned, the status of the vehicle changes to available, and the customer-vehicle relationship is eliminated so that another person can rent it. The vehicle rental benefits are sent to the digital wallet of the contract owner  who launched it to the Ethereum network. If funds were left over, these would be sent to the client's Ethereum account.

\subsection{Implementation}

\subsubsection{Smart Contract Implementation} 

To deploy the contracts in the blockchain you need to specify the network. To do this, we need to create a configuration file, ’truffle.js’ which is taken into account when compiling and launching the contract.

The following figure shows the connection between the client (browser) and the smart contract hosted on the blockchain. The Truffle connector is used, which allows this connection

\begin{figure}[!th]
\begin{center}
\includegraphics[scale=1.2]{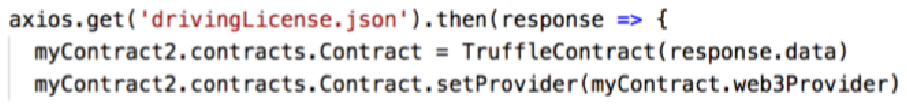}
\caption{Metamask}
\label{fig: Metamask}
\end{center}
\end{figure}

Finally, since Metamask is a browser extension and is the one that interacts with the Ethereum nodes, that is why we interact with the contract from the browser side. To do this, the .json files generated from the contract compilation phase must be loaded in the client side

This section details in deep some relevant functions of the smart contract in charge of the entire rental process of a vehicle. On the one hand, the "Rent a Car" function receives all the parameters of the customer and the vehicle to be rented. This function verifies that the car is not rented and that the license it has received by parameter exists in the other contract

The main function of the smart contract, renting vehicles, is shown in the following figure. This function, written in Solidity, receives the client's data by parameters and links them to the required car

\begin{figure}[!th]
\begin{center}
\includegraphics[scale=1.2]{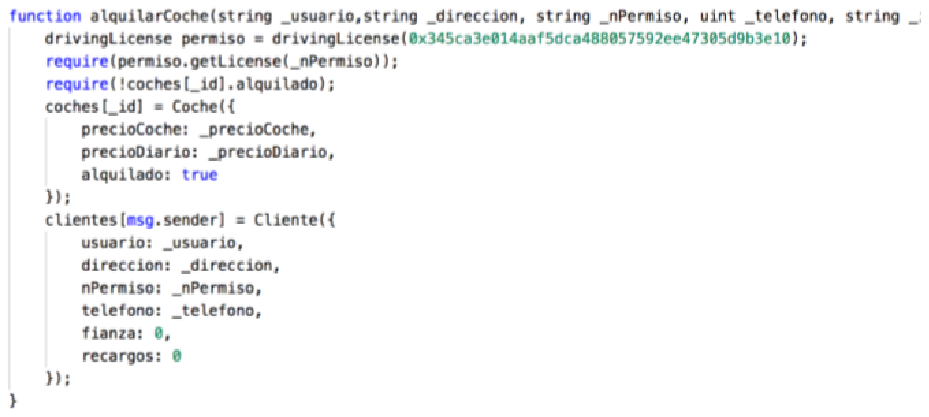}
\caption{Rent a car function}
\label{fig:Rent a car function}
\end{center}
\end{figure}

As for the return function, it verifies that the customer has no pending charges. If the client has no pending charges, the deposit that has been left over to the client is returned. When the car is returned, the owner of the application is given its rental benefits and the car becomes as unrented

\begin{figure}[!thb]
\begin{center}
\includegraphics[scale=1.4]{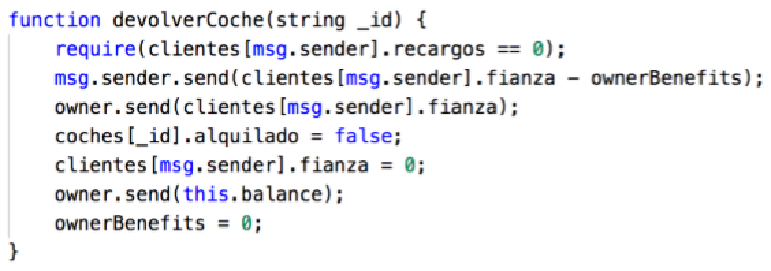}
\caption{Return car function}
\label{fig:Return car function}
\end{center}
\end{figure}

The function written in Solidity checks that the client has no pending charges \textit{(require (clients.charges == 0)}
Send the benefits of the transaction to the owner of the contract \textit{(owner.send)}

\section{Security Analysis Draft}

Users are not registered in any database so their information is not vulnerable to security attacks. The information is managed by the smart contract.
Transactions are not centralized in a Virtual POS, since they are handled by Metamask (Ethereum wallet) so the application does not manage bank information. In addition, blockchain wallets use asymmetric cryptography for transactions increasing the complexity of currency theft attacks.
Blockchain integrity makes it virtually impossible to maliciously alter any data from a decentralized application that is hosted on it. (Hash check)
Decentralization of the blockchain causes any denial-of-service attack (DDoS) to be computationally very difficult, since throwing a node would not change anything and all nodes but one would have to be thrown to stop being a distributed network, so applications that use this network are less exposed to these attacks.

\section{Conclusions and Future Works}

This work  describes the  design  of a decentralized rental system based on smart contracts and the Ethereum public blockchain.
It also includes the presentation of an implementation  called AutoRent, which was developed following the standards of a DApp to check the performance of the proposed system.
This implementation is open source and no enterprise controls the tokens. Besides, all data of rental car service and customers are stored in a public and decentralized blockchain. Finally, the application uses Ethers as cryptographic tokens, the PoW protocol for the transactions.

The guidelines of this work imply different characteristics from traditional applications web. The user does not need to register, does not store password and there is no control over the user’s sessions. The Smart Contract is in charge to use customer’s digital wallet and store the data in the same one. The Smart Contract gives to the application autonomy because it is in charge of the renting and returning of the vehicles, storing and returning the money, distributing and charging automatically to each customer every day, avoiding other ways of payment.
Internet of Things, Artificial Intelligence and blockchain will definitely settle in the applications development in the next years. Will be smart applications with autonomy and capacity to manage and take decisions autonomously.

Among possible future works, we highlight the following. A public blockchain can be used to store all data of the citizens (Driving license, ID, location, etc.). Any decentralized application will be able to use the data of this blockchain for its smart contracts and most of these registries and traditional databases will disappear because the user’s authentication and verification will be made in the public contracts. 

In the very near future, with the wide deployment and development of the Internet of Things and Artificial Intelligence, all rental car networks will have geolocation sensors to know the exact location of each vehicle in every moment. Besides, vehicles will have digital keys that will be transferred by the contract in the transaction moment. Thus, customers will not need going to a physical point to take the keys. This will allow that  customers can left the car where ever they want and the next customer can go to those point because the car is geolocated.
Furthermore, the fluctuation in the value of the cryptocurrency can recommend the consideration of issuing an own token with a stable value.
Another aspect that deserves a study is a study of the General Data Protection Regulation (GDPR) and the impact it will have on the blockchain, both in public and privates in order to make changes in the proposal if necessary to comply it.

\section*{Acknowledgment}
Research supported by the Spanish Ministry of Science, Innovation y Universities and the Centre for the Development of Industrial Technology CDTI  under Projects RTI2018-097263-B-I00 and C2017/3-9.

\end{document}